\documentclass[fleqn,twoside]{article}
\usepackage{espcrc2}

\usepackage{graphicx}
\usepackage[figuresright]{rotating}

\newcommand{\sh}{\slash\!\!\!\!}
\newcommand{\bbox}[1]{\mbox{\boldmath $#1$}}
\newcommand{\case}[2]{{\textstyle\frac{#1}{#2}}}
\newcommand{\gd}{\bbox{\gamma}\!\cdot\!\bbox{D}}
\newcommand{\de}{\bbox{D}\!\cdot\!\bbox{E}}
\newcommand{\ed}{\bbox{E}\!\cdot\!\bbox{D}}
\newcommand{\db}{\bbox{D}\!\times\!\bbox{B}}
\newcommand{\bd}{\bbox{B}\!\times\!\bbox{D}}
\newcommand{\ale}{\bbox{\alpha}\!\cdot\!\bbox{E}}
\newcommand{\sib}{\bbox{\Sigma}\!\cdot\!\bbox{B}}
\newcommand{\vp}{\bbox{p}}

\newcommand{\AmS}{{\protect\the\textfont2
  A\kern-.1667em\lower.5ex\hbox{M}\kern-.125emS}}

\title{A Relativistic $O(a^2)$ Improved Action for Heavy Quarks}
\author{M. B. Oktay\address[MCSD]{Department of Physics, University of 
        Illinois, 
        1110 West Green Street, Urbana, IL 61801}\thanks{Talk presented by
	M. B. Oktay.},
        A. X. El-Khadra{\addressmark \address{Fermi National
	Accelerator Laboratory, 
        P.O. Box 500, Batavia, IL 60510}}, 
        A. S. Kronfeld\addressmark,
        P. B. Mackenzie \addressmark,  and J. N. Simone\addressmark}

\begin{document}

\begin{abstract}
We extend the Fermilab formalism for heavy quarks to develop an
$O(a^2)$ improved relativistic action.
We discuss our construction of the action, including the identification of
redundant operators and the calculation of the improvement coefficients.
\vspace{1pc}
\end{abstract}

\maketitle

\section{INTRODUCTION}

A major source of uncertainty in numerical simulations of lattice QCD
comes from finite lattice spacing effects. Since these effects arise at 
short  distances, they can be analyzed using an effective field theory,
as first proposed by Symanzik~\cite{sym1}.
We may write the effective Lagrangian (LE$\mathcal{L}$) as
\begin{equation}
{\mathcal{L}}^{\mathrm{eff}}={\mathcal{L}}^{\mathrm{cont}}+\sum_j c_j a^{s_j-4}
O_j,
\label{LEL}
\end{equation}
where $s_j=\mathrm{dim}[O_j]$.
In this framework lattice spacing effects can be systematically removed 
by adding higher dimensional operators to the lattice action.
The aim of this work is to design a lattice action such that the coefficients
of $O(a)$ and $O(a^2)$ terms in Eq.~(\ref{LEL}) can be reduced.

For heavy quarks with 
$m_Q\gg\Lambda_{\mathrm{QCD}}$,~the  
mass introduces an additional short-distance scale into the problem,
and the Symanzik formalism must be modified to separate the 
short-distance effects of both the lattice spacing and the heavy-quark 
mass from the long-distance physics. The Fermilab formalism \cite{kkm}
represents one approach to this problem. 
It allows the coefficients $c_j$ in Eq.~(\ref{LEL}) to depend on the quark mass.
Then it takes the
Wilson action~\cite{wilson}, modified to allow different coefficients 
for space-like and time-like operators. Improved lattice actions
are constructed by adding higher-dimension operators, again with different
coefficients for time- and space-like operators. Ref.~\cite{kkm} considers
operators up to dimension five, to obtain a lattice action for fermions,
improved to $O(a)$, and valid for quarks with 
arbitrary mass. In this work we extend the analysis of Ref.~\cite{kkm} 
to include operators of dimension six for $O(a^2)$ improvement.

\begin{table}[htb]
\caption{Dimension-six bilinear interactions that could 
appear in the effective Lagrangian.}
\label{table:1}
\renewcommand{\tabcolsep}{0.925pc} 
\renewcommand{\arraystretch}{1.3} 
\begin{tabular}{@{}rlc}
\hline
No  & Operator  &  Parameter \\
\hline
1  &  $\bar\psi(\gamma_0D_0)^3\psi$ & $\epsilon_2+\overline\epsilon_2$ \\
2 &  $\bar\psi\{\gd,\bbox{D}^2\}\psi$ \\
3 &  $\bar\psi\{\gamma_0D_0,(\gd)^2\}\psi$ & $\delta_2+\overline\delta_2$ \\ 
4 & $\bar\psi[\gamma_0D_0,(\gd)^2]\psi$ & $\delta_2-\overline\delta_2$ \\
5 & $\bar\psi\{(\gamma_0D_0)^2,\gd\}\psi$ & $\delta_u-\overline\delta_u$ \\ 
6 & $\bar\psi[(\gamma_0D_0)^2,\gd]\psi$  & $\delta_u+\overline\delta_u$ \\
7 &$\bar\psi\bbox{\gamma}\!\cdot\![D_0,\bbox{E}]\psi$&$\epsilon_F+\overline\epsilon_F$\\
8 &  $\bar\psi[\ale,\gamma_0D_0]\psi$ & $\epsilon_F-\overline\epsilon_F$ \\
9 & $\bar\psi\{\gd,\ale\}\psi$ \\
10 &  $\bar\psi[\ale,\gd]\psi$ & * \\
11 &  $\bar\psi\{i\sib,\gamma_0D_0\}\psi$ &$\delta_B+\overline\delta_B$ \\
12  & $\bar\psi\{i\sib,\gd\}\psi$  \\
13 & $\bar\psi[i\sib,\gd]\psi$  \\
14 & $\bar\psi[i\sib,\gamma_0D_0]\psi$ & $\delta_B-\overline\delta_B$ \\
15 & $\bar\psi\gamma_0(\de-\ed)\psi$   \\
16 & $\bar\psi\bbox{\gamma}\!\cdot\!(\db+\bd)\psi$ \\
17 &  $\bar\psi(\gamma_iD_i^3)\psi$  \\
\hline
\end{tabular}
\end{table}

\section{THE ACTION}

At dimension six we have to consider both fermion bilinears and 
four-fermion operators. For these proceedings, we concentrate on 
determining the bilinear interactions. 
The five dimension-six bilinears that satisfy axis-interchange symmetry 
are given in Ref.~\cite{sw}. 
Without axis-interchange symmetry, these operators become
the seventeen operators listed in Table~\ref{table:1}. 
(The lower dimensional operators are given in Ref.~\cite{kkm}.) 

Following Refs.~\cite{sw,kkm}, we use field transformations to
expose the redundant directions for on-shell improvement.
Since we are considering dimension-six operators in the
action we must include dimension-two operators in the transformations,
and we must also consider gauge field transformations.
Writing
\begin{equation}
\psi\rightarrow e^{J}\psi, \quad \bar\psi\rightarrow \bar\psi e^{\bar J},
\label{eq:fermi}
\end{equation}
we have
\begin{eqnarray}
J&=&a\epsilon_1(\sh D+m)+a\delta_1\gd +a^2\epsilon_2(\sh D+m)^2
\nonumber \\
&-&a^2\epsilon_F\case{i}{2}\sigma_{\mu\nu}F_{\mu\nu}+a^2\delta_2(\gd)^2
+a^2\delta_Bi\sib \nonumber \\
&+&a^2\delta_u[\gamma_0D_0,\gd],
\end{eqnarray}
and $\bar J$ is the same but with bars over the parameters. For the gauge
fields 
\begin{eqnarray}
A_0 \to A_0 & + & \case{1}{2}a^2\epsilon_A(\de-\ed),
\label{eq:gauge1} \\
	& + & g_0^2a^2\epsilon_J(\bar\psi\gamma_0t^a\psi)t^a
\nonumber \\
\bbox{A} \to \bbox{A}  & + & \case{1}{2}a^2(\epsilon_A+\delta_A) (\db+\bd)
 \nonumber \\
& - & \case{1}{2}a^2(\epsilon_A+\delta_E)[D_0,\bbox{E}] 
\label{eq:gauge2} \\
& + & g_0^2a^2(\epsilon_J\!+\delta_J)(\bar\psi\bbox{\gamma}t^a\psi)t^a.
\nonumber
\end{eqnarray}
Following the notation of  Ref.~\cite{kkm}, the $\epsilon$ ($\delta$) 
coefficients label axis-interchange symmetric (asymmetric) operators, 
so that the axis-interchange symmetric analysis of Ref.~\cite{sw} 
is recovered when all $\delta$ coefficients vanish.
The transformations of Eqs.~(\ref{eq:fermi})--(\ref{eq:gauge2}) 
generate the operators listed in 
Table~\ref{table:1} with
coefficients which depend on the parameters in the transformations.
The last operator of Table~\ref{table:1}, $\bar\psi\gamma_iD_i^3\psi$,
is not generated by the transformations, and hence cannot be eliminated.
To remove it one must improve the nearest-neighbor lattice derivative.

Not all the operators are independent from each other. Indeed, we have
used the identities 
\begin{eqnarray}
2\gamma_0D_0\gd\gamma_0D_0 & = & \label{eq:rel1} \\
-\bbox{\gamma}\!\cdot\![D_0,\bbox{E}] & + & \{(\gamma_0D_0)^2,\gd\},
\nonumber \\
2\gd\gamma_0D_0\gd & = & \label{eq:rel2} \\
\{\gd,\ale\} & - & \{\gamma_0D_0,(\gd)^2\},
\nonumber 
\end{eqnarray}
\begin{equation}
2(\gd)^3 = \{\gd,\bbox{D}^2\} + \{\gd,i\sib\},
\label{eq:rel3}
\end{equation}
to remove the operators shown on the left hand side of these equations 
from Table~\ref{table:1}.
The interaction 
$\bar\psi[\ale,\gd]\psi$ (marked ``*'' in Table~\ref{table:1}) does not
obey particle-antiparticle interchange symmetry so it  
cannot be in the action.
Nine independent parameter combinations appear in the coefficients of
the transformed operators. They are
$\epsilon_2+\overline\epsilon_2$,
$\delta_2+\overline\delta_2$, $\delta_2-\overline\delta_2$, 
$\delta_u+\overline\delta_u$, $\delta_u-\overline\delta_u$, $\delta_B+\overline\delta_B$, $\delta_B-\overline\delta_B$, $\epsilon_F+\overline\epsilon_F$ and 
$\epsilon_F-\overline\epsilon_F$. 

The choice of redundant operators is not unique. Considerations,
such as calculational convenience (or solving the fermion doubling
problem) play a role in the choice. In this work, we want to avoid
operators which contain higher order time derivatives, as they spoil
the good properties of the transfer matrix of actions with Wilson-like
time derivatives. There are nine such operators, all
of which can be eliminated by the field transformations. 
Table~\ref{table:1} shows which parameter combination is used to
eliminate a given operator. In summary, this analysis reduces the original 
seventeen operators to seven, and we can write the lattice fermion action as
\begin{eqnarray}
\lefteqn{S_{\mathrm{F}}=S_0+S_B+S_E+a^2c_1\int\bar\psi\{\gd,\bbox{D}^2\}
\psi}  \nonumber \\
\lefteqn{+a^2c_2\int\bar\psi\gamma_iD_i^3\psi+a^3c_3\int\bar\psi\{i\sib,
\gd\}\psi} \nonumber \\
\lefteqn{+a^2c_4\int \bar\psi\gamma_0(\de-\ed)\psi}
\label{eq:action} \\
\lefteqn{+a^2c_5\int\bar\psi\bbox{\gamma}\cdot(\db+\bd)\psi}
\nonumber \\
\lefteqn{+a^2c_6\!\int\!\!\bar\psi\{\gd,\ale\}\psi+a^2c_7\!\!\int\!\!
\bar\psi[i\sib,\gd]\psi.}
\nonumber
\end{eqnarray}
\normalsize
where $S_0$, $S_B$ and $S_E$ are given in Ref.~\cite{kkm}.

\section{THE COEFFICIENTS}

Conditions on the coefficients of the improvement operators are
obtained by matching on-shell quantities in the lattice theory to
their continuum counterparts. From the lattice fermion propagator we
derive the dispersion relation
\begin{eqnarray}
\cosh E=1+{(\mu(\vp)-1)^2 +(f(\vp)\bbox{S}+2c_2\bbox{Q})^2
\over 2{\bf\mu(\vp)}}
\end{eqnarray}
where (in lattice units) 
\begin{eqnarray}
\mu(\vp)&=&1+m_0+\case{1}{2}r_s\zeta\hat{\bbox{p}}^2 \\
f(\vp)&=&\zeta-2c_2-2c_1\hat{\bbox{p}}^2
\end{eqnarray}
$S_i=\!\sin p_i$, 
$\hat{p}_i=2\sin(p_i/2)$ and $2Q_i=\sin 2p_i$. 
Expanding the energy in powers of $\bbox{p}$, we obtain
\begin{eqnarray}
E=M_1+{\vp^2\over 2M_2}-{1\over 6}w_4\sum_{i=1}^3 p_i^4-
{(\vp^2)^2\over 8M_4^3}+ \dots
\end{eqnarray} 
where $M_1$ is the rest mass and $M_2$ is the kinetic mass. 
As in Ref.~\cite{kkm} we impose $M_2=m_q$.
For a relativistic LE${\cal L}$ we also impose $M_1=M_2$.
The improvement condition 
$M_2=M_4$ yields a relation for the coefficient $c_1$.
Rotational invariance ($w_4=0$) gives a condition for $c_2$. 
These relations were already derived in Ref.~\cite{kkm}.

For the remaining operators, $c_3$ through $c_7$, we calculate temporal and 
spatial matrix elements with one gluon exchange at tree-level.
Using the lattice spinors of Ref.~\cite{kkm}, we expand the temporal 
matrix element up to and including $O(p^2)$. The lattice
matrix element is
\begin{eqnarray}
\lefteqn{
\langle q(\xi',\vp')|V_0^{\mathrm{G}}|q(\xi,\vp)\rangle_{\mathrm{lat}}
=u^\dagger(\xi',{\bf\vec{0}})\Big[1- } \nonumber \\
\lefteqn{\hspace*{1em}
{(\vp'{}^2+\vp^2-2\vp'.\vp)\over 8M_E^2}
+{i\epsilon_{ijl}\Sigma_l p'_ip_j\over 4M_{E'}^2}\Big]u(\xi,{\bf\vec{0}}),}
\label{eq:VG0}
\end{eqnarray}
where
\begin{eqnarray}
\lefteqn{\frac{1}{8M_E^2} = {\zeta^2\over 2m_0^2(2+m_0)^2}+{c_E\zeta^2\over 2m_0(2+m_0)}} \nonumber \\
\lefteqn{+{c_6-c_4\over (1+m_0)},} \\
\lefteqn{\frac{1}{4M_{E'}^2}=\frac{1}{4M_E^2}+{2c_4\over 1+m_0}.}
\end{eqnarray}
The right-hand side of Eq.~(\ref{eq:VG0}) must be matched to the
continuum matrix element, which has $M_E$ and $M_{E'}$ replaced
with~$m_q$.
One matching condition is to set $M_{E'}=M_E$, which requires $c_4=0$.
Another matching condition sets $M_E=M_2$, yielding a condition on
$c_6$ and $c_E$:
\begin{eqnarray}
\lefteqn{{4\zeta^2\over m_0^2(2+m_0)^2}\!+\!{4c_E\zeta^2\over m_0(2+m_0)}\!+\!
{8c_6\over 1+m_0}\!\!=\!\!{1\over M_2^2}.}
\label{eq:cec8}
\end{eqnarray}
Eq.~(\ref{eq:cec8}) is in agreement with a result from 
Ref.~\cite{kkm} obtained 
in the Hamiltonian formalism. 
(The expression for $M_2$ and the definition of $c_E$ can be found in
Ref.~\cite{kkm}.)

The determination of improvement conditions on $c_3$, $c_5$ and $c_7$,
requires the calculation of
the spatial matrix element including terms up to $O(p^3)$.
This is currently in progress.

\section{CONCLUSIONS}

We propose a relativistic $O(a^2)$ improved action for heavy 
quarks. We use the redundant directions to eliminate all operators with 
higher order time derivatives. As a result, our action keeps the Wilson
time derivative. We have determined the mass dependent coefficients of 
four improvement operators at tree-level. The 
determination of all remaining coefficients
is in progress.

\vskip 8pt
M.B.O.~and A.X.K.~are supported in part by the U.S.\ Department of
Energy under contract DE-FG02-91ER40677, and
Fermilab is operated by Universities Research Association Inc.,
under contract with the DOE.

\end{document}